\documentclass[letterpaper,english]{elsart}
\usepackage[T1]{fontenc}
\usepackage[latin1]{inputenc}
\usepackage{graphicx}

\makeatletter

\usepackage{subfigure}
\usepackage[percent]{overpic}
\usepackage[usenames]{color}

\begin{document}
\begin{frontmatter}

\title{Arrays of widely spaced atomic steps on Si(111) mesas due to sublimation}

\author{Kee-Chul Chang}
\ead{kc53@cornell}
\author{Jack M. Blakely}
\address{Department of Materials Science and Engineering, Cornell University,
Ithaca, NY 14853}

\begin{abstract}
Steps with spacings of microns form on top of mesas fabricated on
Si(111) that is annealed at temperatures where sublimation becomes
important. Upon annealing, mesas first develop ridges along their
edges, effectively creating craters which then become step-free by
a step flow process described in the literature\cite{Tanaka:1996,Homma:1996}.
Due to the miscut of the average surface from (111), ridge breakdown
occurs on one edge of each mesa as sublimation proceeds. The breakdown
point then acts as a source of steps which spread out over the mesa
surface. The distribution of steps in the resulting step train depends
on the sublimation rate, direct step-step interaction and the diffusive
exchange of atoms among the steps. Insight into the role of these processes on the self-organization of the wide
terrace distributions is provided by computer simulations using BCF
(Burton, Cabrera and Frank) theory.  This shows that step spacing can be controlled
by varying the annealing temperature and the deposition flux.
Comparison of the experimental and predicted step distributions suggest that the dynamics 
of the widely spaced steps are dominated by sublimation.
\end{abstract}

\end{frontmatter}


\section{Introduction}

Atomic steps on crystal surfaces are the basic units of surface morphology;
by controlling the local density of steps, we can control the surface
shape. The contour and density of surface steps can be manipulated
by a combination of patterning, which defines the initial step morphology,
and thermal annealing, which results in the movement and creation/destruction
of steps. The ability to guide step dynamics through these processes
represents the ultimate limit in height morphology control.

There are many useful applications for surfaces with controlled step
morphologies. Due to the changes in coordination at atomic steps,
adsorption will usually occur preferentially at those locations. Steps
are known to have extremely important effects in nucleation and continued
growth of homo- and hetero-epitaxial layers.

Previous work in this field has focused on the creation of large step-free
areas. Step-free areas 10-100$\mu$m in radius have been formed
on Si\cite{Lee:2000,Homma:1996,Tanaka:1996}, SiC\cite{Powell:2000}
and sapphire\cite{Chang:2002} surfaces by several groups, including
our own; these may be useful for making devices that require high
quality interfaces\cite{Oliver:2000} as well as providing model surfaces to investigate
nucleation and defect properties\cite{Tromp:1998}. But arrays of
steps that have controlled spacing in the micron range may also have
use in applications such as templates for nanowires or organic thin
films. Their study may also lead to better models of step dynamics
for weakly interacting steps. 

Widely spaced steps are difficult to make because orientational errors
of 0.1$^{\circ}$ or more are usually introduced in aligning crystals
during polishing, resulting in vicinal surfaces with terrace widths
in the 100nm range on Si(111). To successfully create such
arrays, an understanding of step dynamics is required to effectively
redistribute the steps. For Si(111), previous literature has focused
on understanding the role of electromigration\cite{Latyshev:1989}
and the Ehrlich-Schwoebel\cite{Ehrlich:1966,Schwoebel:1966} effect
in creating step bunches.

In this Paper, we describe an experiment in which mesa structures
(areas that are completely surrounded by trenches) on Si(111) were
annealed at temperatures over 925$^{\circ}$C in ultrahigh vacuum to
produce arrays of steps with spacing of 1$\mu$m or more. In our discussion, we show that 
these arrays arise through transient ridge formation around the edges of
the mesas followed by redistribution of the steps on top of the mesa
after ridge breakdown.

\section{Experimental}

Si(111) was patterned by standard optical lithographic methods using
silicon oxide as an etch mask. Mesa structures with different azimuthal
orientations and size were etched into silicon by reactive ion etching.
The etch depth was around 2$\mu$m 
for all of our samples. 15mm$\times$5mm samples were cut from the
wafer using a diamond tip scriber. The samples were dipped in HF for
3 minutes to remove any residual oxide and to hydrogen terminate the
surface before being loaded into the vacuum chamber.

The samples were heated by direct current in a UHV system (base pressure
$10^{-10}$ Torr) while the temperature was monitored with a pyrometer.
Some samples were subjected to flashing above 1200$^{\circ}$C to
quickly remove the native oxide layer before annealing while others
were annealed without flashing. The annealing temperature ranged from
925$^{\circ}$C to 1150$^{\circ}$C. After annealing, the samples were
taken out of the vacuum chamber and imaged by AFM in air.

\section{Results \& Discussion}

\begin{figure}
\subfigure{
\label{fig:mesa_ridge}
  \begin{overpic}[%
  width=5in,
  keepaspectratio]{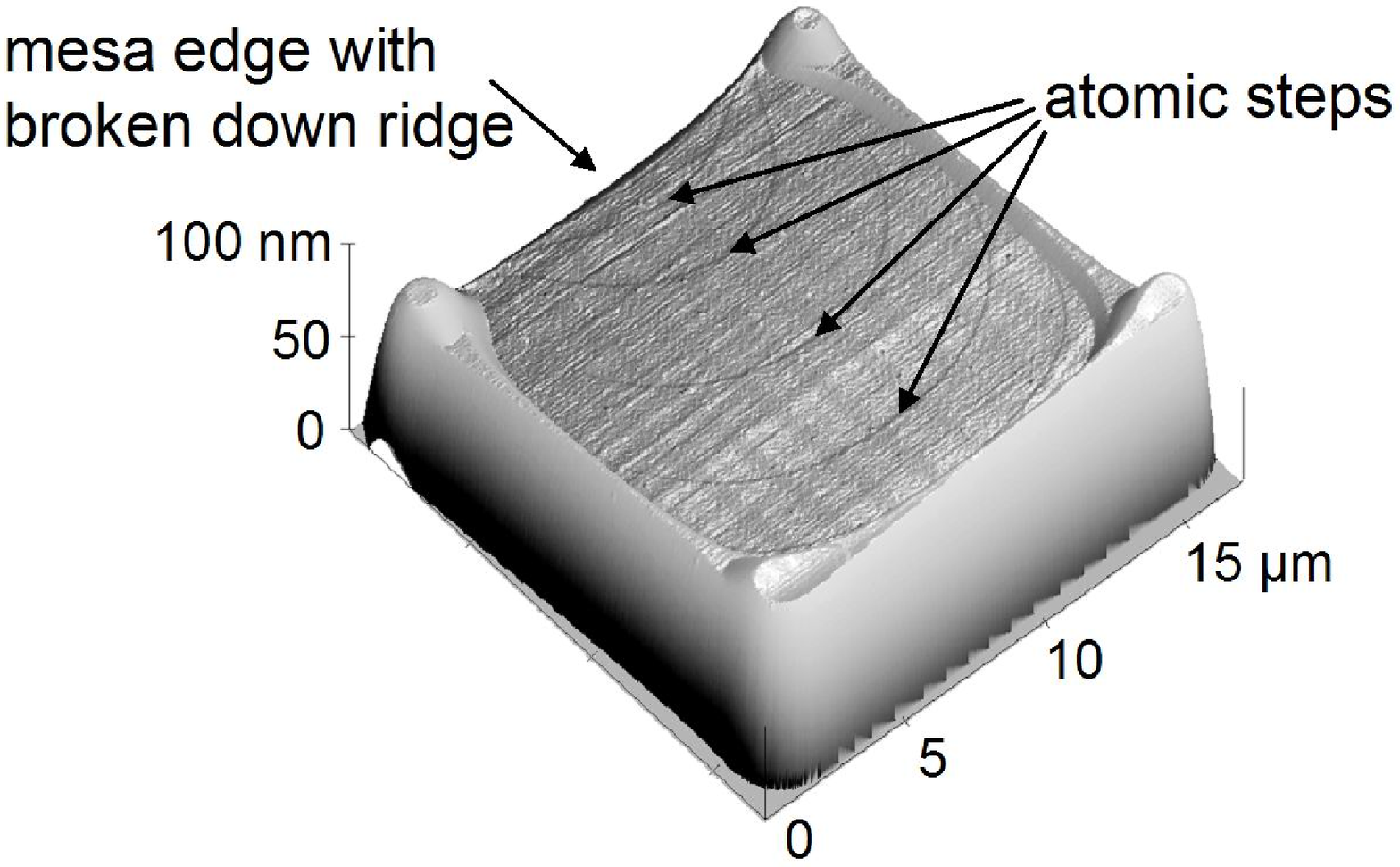}
  \put(0,0){\textbf{\subref{fig:mesa_ridge}}}
  \end{overpic}
}
\subfigure{
\label{fig:step_array}
  \begin{overpic}[%
  width=5in,
  keepaspectratio]{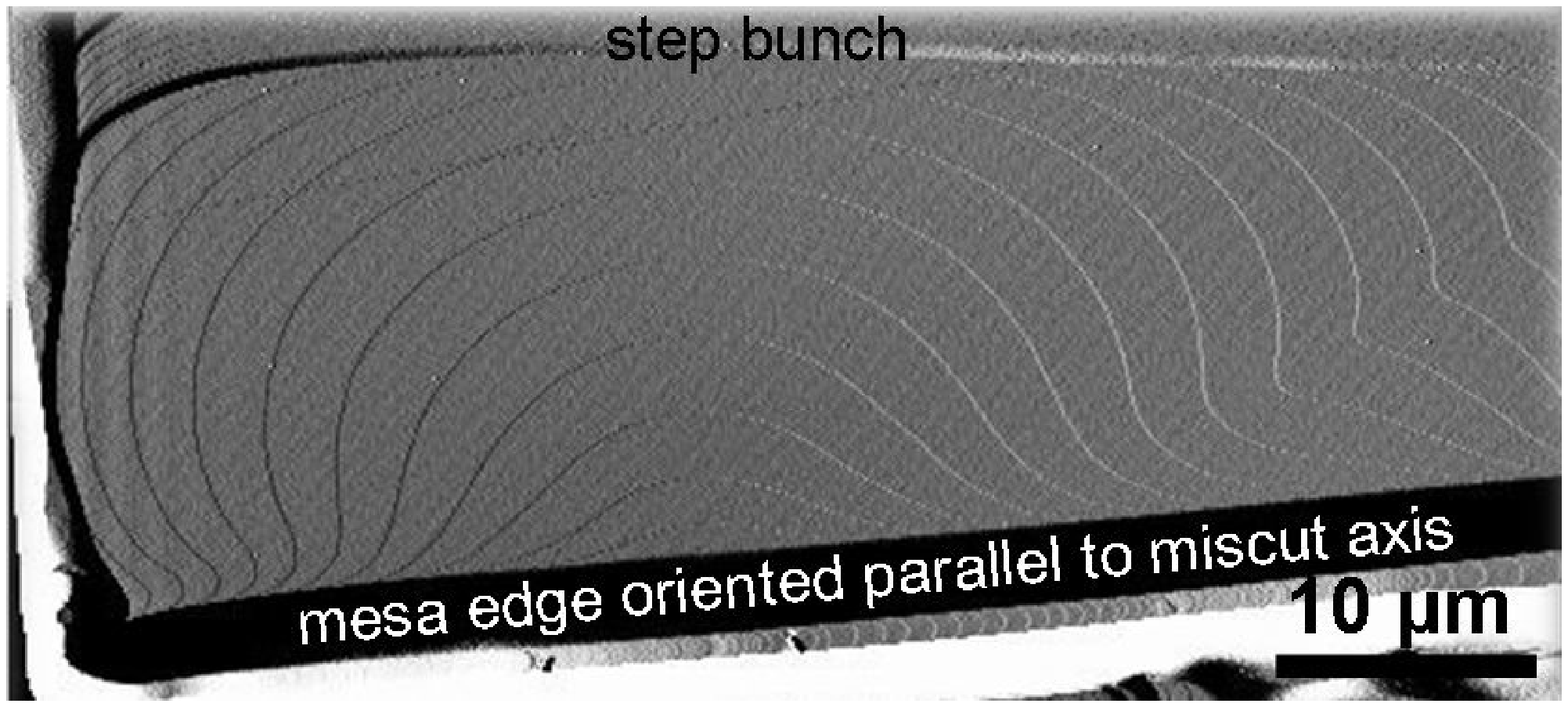}
  \put(0,0){\colorbox{White}{\textbf{\subref{fig:step_array}}}}
  \end{overpic}
}

\caption{\subref{fig:mesa_ridge} Height profile of a mesa on Si(111) that
has been annealed at 1000$^{\circ}$C for 1 hour followed by flashing
at 1220$^{\circ}$C for 20 seconds.  Note that a ridge exists along three
edges of the mesa. \subref{fig:step_array} Deflection AFM image of step
arrays formed on top of a rectangular mesa on Si(111) after flashing
at 1250$^{\circ}$C followed by annealing at 1100$^{\circ}$C for 15 minutes. }
\end{figure}

Fig.~\ref{fig:mesa_ridge} shows an example of the typical height
profile of mesas after annealing. Ridges have formed along most of
the edges of the mesas. The portion of the edge without a ridge 
act as a step source for the widely separated step arrays that
develop on the mesa tops.  The step arrays formed through this process
showed similar step spacing on other mesas on the sample that were of the same size and orientation.

From a series of rotated mesas patterned on the sample, we also find that
the length of the mesa edge without a ridge increases dramatically
as the mesa edge becomes nearly parallel to the miscut axis as illustrated
in Fig.~\ref{fig:step_array}. The crystallographic directions
of the mesa edges do not affect the amount of ridge breakdown. The
direction of the DC heating current (and hence electromigration effects)
also appears to be unimportant, and in fact similar results were obtained
with low frequency (0.2 Hz) AC heating.

We will look at the two stages in the formation of the step arrays viz (i) the development and 
breakdown of ridges along the mesa edges and (ii) the generation of widely spaced steps across the mesas.

\subsection{Spontaneous Ridge Formation on Mesa Edges}

The formation of ridges along the edges of mesas is predicted\cite{Liau:1990} by the
classical Mullins theory\cite{Mullins:1957,Mullins:1959} based on
the tendency of the material to eliminate chemical potential gradients
associated with high curvature at the mesa edges.  However, Mullins theory
is expected to be applicable above the roughening temperature whereas
our experiments were performed below the roughening temperature of the Si(111) surface.
When below the roughening temperature, the elimination of the sharp edges requires
mass flow to the mesa top that results in either the formation of new islands or
incorporation of atoms into pre-existing steps.\footnote{For this to occur, 
the roughening temperature only needs to be exceeded for  
a range of surface orientations between the mesa top and the sidewall.}

Mullins, Rohrer\cite{Mullins:2000,Rohrer:2001}
and Combe\cite{Combe:2000} have considered the case where ridge formation requires that
new islands have to be nucleated on the mesa surface; they conclude that once
the radius of curvature of the edge exceeds $\sim$1nm, the free energy
barrier for further island nucleation process becomes prohibitively large. 

However, the existence of miscut steps on the mesa top can significantly reduce
the barrier for ridge formation if the transfer of atoms occurs to these existing steps; we believe
this to be the most likely mechanism on these patterned surfaces.  Thus ridges can be produced
by modification of the shape of the miscut steps on the mesa top. 

The formation of ridges by this process is qualitatively different from that predicted by
the classical Mullins theory because the ridge height will be dependent on the number of
miscut steps near the mesa edge.  This is in line with our observation of ridges on the mesa
that is aligned with the miscut steps.  Near the lower edge, a ridge of lower height forms and can
subsequently disappear as sublimation proceeds.

\begin{figure}
\subfigure{
\label{fig:sim_ridge}
\begin{overpic}[%
  width=2.5in,
  keepaspectratio]{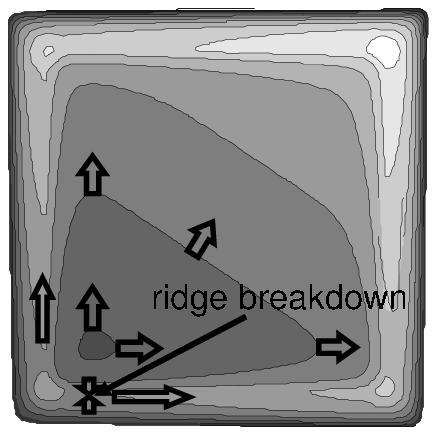}
  \put(3,90){\textcolor{White}{\textbf{\subref{fig:sim_ridge}}}}
\end{overpic}
}
\subfigure{
\label{fig:circular}
\begin{overpic}[%
  width=2.5in,
  keepaspectratio]{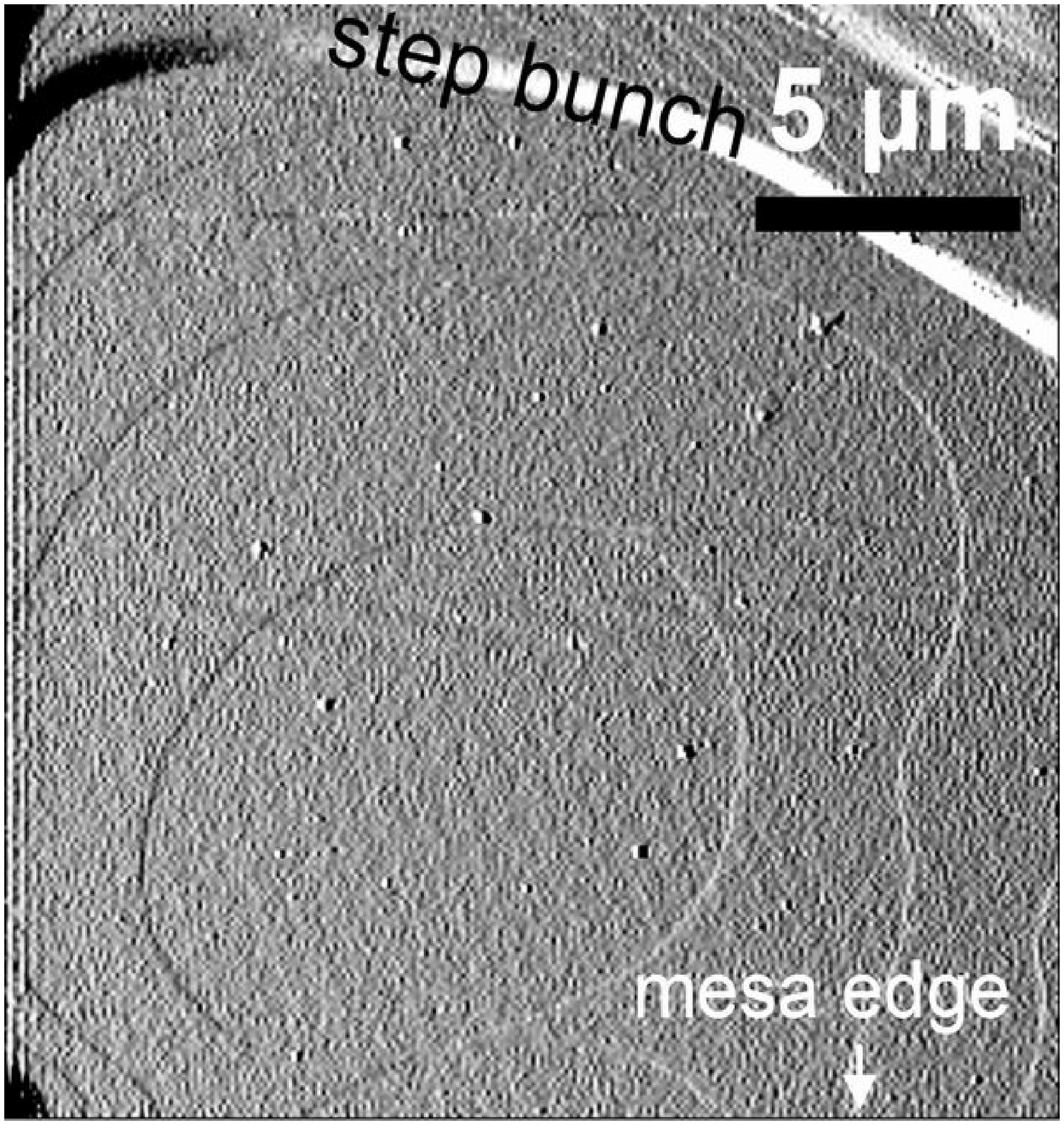}
  \put(3,90){\textcolor{White}{\textbf{\subref{fig:circular}}}}
\end{overpic}
}

\caption{\subref{fig:sim_ridge} Step motion on top of the mesa after
mesa edges have rounded enough to slow diffusion-induced ridge formation. Ridges convert 
the centers of the mesa into craters and the steps forming these craters expand outwards to create
a central step-free area. \subref{fig:circular} Steps near an edge of a mesa on Si(111) after
flashing at 1400$^{\circ}$C and quenching. The circular steps are similar to those previously
reported\cite{Lee:2000,Homma:1996} for craters surrounded by patterned ridges.}
\end{figure}

The mass flow from the edges and the corners of the mesa that creates
ridges will slow down as the radius of curvature of the edges increase. Such
a structure is illustrated in Fig.~\ref{fig:sim_ridge}.
During sublimation, the steps forming the ridges will retract toward
a corner of the mesa, resulting in configurations of steps of both
signs. The steps on top of the ridge will eventually meet the opposite
sign step on the other side of the ridge, resulting in a reduction of
its height. Repetition of this process will result in ridge breakdown.

While the ridge is evolving, the slight depression on
the mesa, formed by ridge enclosure, begins to become step-free
due to sublimation. When the ridge finally breaks down, a low
step density region, bounded by step bunches, is formed. A possible snapshot
of this process on a sample that had been flashed to 1400$^{\circ}$C
and quickly quenched is shown in Fig.~\ref{fig:circular}.

\subsection{Development of Widely Spaced Step Arrays on Mesa Surfaces}

\begin{figure}
\subfigure{
\label{fig:csim_setup}
\begin{overpic}[%
  width=4in,
  keepaspectratio]{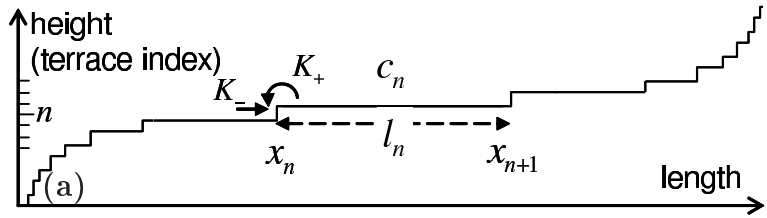}
  \put(5,3.5){\textbf{\subref{fig:csim_setup}}}
\end{overpic}
}
\subfigure{
\label{fig:adl_sl}
\begin{overpic}[%
  width=4in,
  keepaspectratio]{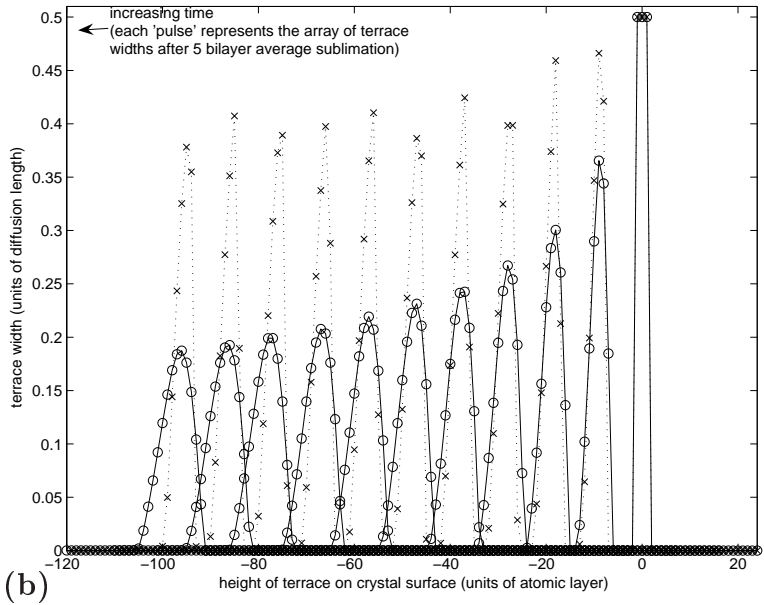}
  \put(0,0){\textbf{\subref{fig:adl_sl}}}
\end{overpic}
}
\subfigure{
\label{fig:ren_tw}
\begin{overpic}[%
  width=4in,
  keepaspectratio]{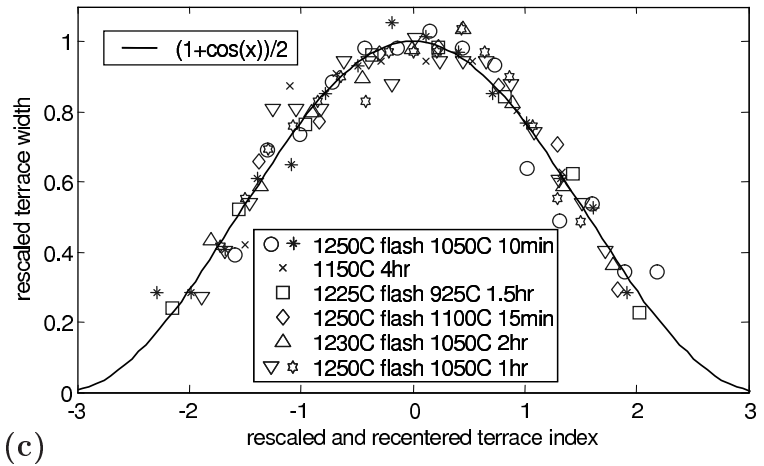}
  \put(0,0){\textbf{\subref{fig:ren_tw}}}
\end{overpic}
}

\caption{\subref{fig:csim_setup} Diagram of the model for computer simulations.
The terrace distribution of a region of low step density bounded by
step bunches as shown is investigated. \subref{fig:adl_sl} Computer simulation results for
the evolution of the terrace distribution for the initial configuration
of 3 terraces 0.5$x_{s}$ wide bounded by step bunches with $10^{-4}x_{s}$ step spacing when $d_{+}=10^{-6}$,
$d_{-}=0$ (solid line) and $d_{+}=0.1$, $d_{-}=0.05$ (dotted line)
plotted in increments of the time to evaporate 5 bilayers. \subref{fig:ren_tw} Experimental
data showing rescaled terrace index (recentered at the maximum terrace
width) versus terrace width fitted to a cosine curve for 8 mesas on
6 samples annealed under different conditions. This suggests that
the terrace length scale for the widely spaced steps in our experiments
is in the sublimation limited regime.}
\end{figure}

After ridge breakdown, widely spaced step arrays form on the mesas and persist even after 
long annealing times in which hundreds of atomic layers of the crystal are removed%
\footnote{The highest annealing temperature of 1100$^{\circ}$C results in the
removal of a bilayer every 5 seconds. We have done anneals
for 2 hours at this temperature and still get widely spaced step arrays.}.  
To better understand the development of these arrays, computer simulation based on the Burton-Cabrera-Frank theory\cite{Burton:1951}
were used to model the evolution of widely spaced steps bounded
by step bunches as illustrated in Fig.~\ref{fig:csim_setup}. Our
system was assumed to be composed only of steps with the same sign,
so that step annihilation does not occur; the 2-dimensional step arrays
were approximated as 1-dimensional, which should be a reasonable approximation for the
steps moving in from the middle of the gap since the radius of curvature
of those steps is so large. In the BCF theory with the quasi-static
approximation, the adatom concentration $c$, on the terraces satisfies
the equation
\begin{equation}
D\frac{\partial^{2}c}{\partial x^{2}}-\frac{c}{\tau}+F=0
\label{eq:bcfeq}
\end{equation}
where $D$ is the diffusion constant of the adatoms, $1/\tau$ is
the rate of evaporation and $F$ is the flux of atoms condensing on
the surface per unit time. (For the experiment reported here $F=0$.) The boundary
conditions for step $n$ at $x_{n}$ is given by
\begin{eqnarray}
D\left.\frac{\partial c_{n}}{\partial x}\right|_{x_{n}} & = & +K_{+}\left(c_{n}(x_{n})-c_{n}^{eq}\right)\\
D\left.\frac{\partial c_{n-1}}{\partial x}\right|_{x_{n}} & = & -K_{-}\left(c_{n-1}(x_{n})-c_{n}^{eq}\right)
\end{eqnarray}
where $c_{n}(x)$ is the adatom concentration between steps $n+1$
and $n$, $K_{\pm}$ are the adatom attachment-detachment rate constants
from the upper and lower terraces. Values of $K_{+}$ and $K_{-}$
that are different from each other model the Ehrlich-Schwoebel barrier.
$c_{n}^{eq}$ is the equilibrium concentration of adatoms at step
$n$; $c_{n}^{eq}\approx c_{0}^{eq}(1+\beta\Delta\mu_{n})$ where
$c_{0}^{eq}$ is the equilibrium concentration of adatoms at a single
isolated straight step, $\beta=1/k_{B}T$, and $\Delta\mu_{n}$ is
the free energy change involved in adding an adatom to step $n$ (incorporating
the step-step interactions) from an isolated step.

Solving the set of differential equations for the adatom concentration
with the boundary conditions give the adatom concentration
on the terraces. The gradients of the adatom concentration give the
fluxes of adatoms into the steps from the terraces to the left and
right of the step, determining the step velocity. The result can be
written by first defining the following two functions,
\begin{eqnarray}
\Phi_{n}^{\pm}=\frac{\tanh(\widetilde{l}_{n}/2)+d_{\pm}}{(d_{+}+d_{-})\coth(\widetilde{l}_{n})+(1+d_{-}d_{+})}\\
\Gamma_{n}=\frac{\Delta\mu_{n+1}-\Delta\mu_{n}}{(d_{+}+d_{-})\cosh(\widetilde{l}_{n})+(1+d_{-}d_{+})\sinh(\widetilde{l}_{n})}
\end{eqnarray}
where the lengths are rescaled by the diffusion length $x_{s}=\sqrt{D\tau}$
so that $\widetilde{l}_{n}=(\widetilde{x}_{n+1}-\widetilde{x}_{n})/x_{s}$
is the terrace width between steps $n+1$ and $n$ and $d_{\pm}=(D/K_{\pm})/x_{s}$
is the rescaled attachment-detachment length which represents how
far the adatom must diffuse at the step edges before attaching to
the step. The step chemical potential is defined as $\Delta\mu_{n}=2g\Omega(\widetilde{l}_{n-1}^{-3}-\widetilde{l}_{n}^{-3})/x_{s}^{3}$
where $g$ is the step interaction energy per unit length of step.
The step velocity is then
\begin{equation}
\frac{d\widetilde{x}_{n}}{d\widetilde{t}}=(1-f+\beta\Delta\mu_{n})\left[\Phi_{n}^{-}+\Phi_{n-1}^{+}\right]+\beta\left[\Gamma_{n-1}-\Gamma_{n}\right]
\label{eq:stepvel}
\end{equation}
where $\widetilde{t}=c_{eq}^{0}\Omega t/\tau$ is in units of the
time required to evaporate a monolayer from a crystal in vacuum while
$f=F\tau/c_{eq}^{0}$ is the rescaled deposition flux. 

To easily relate our simulation results to experiment, we will consider
the dynamics in terms of the terrace widths. The rate of change
of a terrace width, $\widetilde{l}_{n}$, can be written as $d\widetilde{l}_{n}/d\widetilde{t}=d\widetilde{x}_{n+1}/d\widetilde{t}-d\widetilde{x}_{n}/d\widetilde{t}$.
This equation was used to simulate the behavior of an array of terrace
widths. Four parameters, $g$, $d_{\pm}$ and $f$ are needed to numerically
integrate this system of differential equations. We set $f=0$ to
model a sublimating surface without any deposition and take $g=5$meV\AA
\cite{Cohen:2002}, $x_{s}=20\mu$m\cite{Homma:1997} from the literature. 

We consider the evolution, as sublimation proceeds, of a set of terraces
that exist between the mesa edge and a step bunch on the mesa. We
take various values of $d_{\pm}$ to investigate the effect of attachment-detachment
rates on the terrace width distribution. Steps move in from the mesa
edge and incorporate into the step bunch on the mesa so periodic boundary
conditions are assumed. The changes in terrace width distribution over
time are displayed in Fig.~\ref{fig:adl_sl} as a series of plots
of terrace index (or terrace height in units of step height) versus
the terrace width. The simulation results in a `pulse' of terrace
widths which move down along the index axis as annealing progresses. 

In the simulation, the pulses seem to self-organize to a stable distribution
over time. With the inclusion of the Ehrlich-Schwoebel effect ($d_{+}>d_{-}$) for
small $d_{\pm}$, the pulse is distributed over a large number of
terraces and decays slowly in amplitude, while for large $d_{\pm}$,
the pulse is formed by a few terraces and quickly becomes stable. By
considering fluctuations on a train of steps, all with the same initial
terrace widths, stable pulse formation was predicted by Misbah\cite{Misbah:1996} and
Sato\cite{Sato:1997} through analysis of the BCF
equation. In the present Paper, the physical argument for pulse formation
is presented for a train of steps that have a wide range of terrace
widths; This allows us to form stable pulses much faster in our computer
simulations.

There are two main physical causes for terrace width redistribution.
For step trains with small terraces, the step motion is driven by
adatom exchange between neighboring steps due to step-step interaction
which tries to equalize the terrace widths of the step train. This
is the case when the second term in Eq.~(\ref{eq:stepvel}) dominates.
The decay rate of fluctuations in terrace widths depends on various
limiting adatom kinetics as discussed in detail by Liu\cite{Liu:1996}.

For step trains with large terraces, the step motion is driven by
sublimation of adatoms from the terraces. This is the case when the
first term in Eq.~(\ref{eq:stepvel}) dominates, and it should be noted that
the terrace widths which result in crossover to this behavior can be controlled
by adjusting the magnitude of the deposition flux, $F$. Sublimation dominated step motion has
sub-regimes depending on the limiting adatom kinetics.

When the terrace widths are small($\widetilde{l}<\sqrt{2(d_{+}+d_{-})}$),
the number of adatoms evaporating from the terraces may be limited
by the attachment-detachment rates at the step. In this attachment-detachment
limited(ADL) regime, the terrace widths in the pulse evolve as described
by the equation\begin{equation}
\frac{d\widetilde{l}_{n}}{d\widetilde{t}}=\frac{(1-f)}{2}\left[\widetilde{l}_{n+1}-\widetilde{l}_{n-1}+\delta(2\widetilde{l}_{n}-\widetilde{l}_{n+1}-\widetilde{l}_{n-1})\right]\label{eq:ADL}\end{equation}
where $\delta=(d_{+}-d_{-})/(d_{+}+d_{-})$. Due to the last term, this equation yields dispersive
traveling wave solutions; fluctuations in terrace width increase in
amplitude with time if $d_{+}>d_{-}$.
This corresponds to the well-known result\cite{Bennema:1973} that
the terrace width distribution is unstable under sublimation with
the Ehrlich-Schwoebel effect. 

For wider terrace widths($\sqrt{2(d_{+}+d_{-})}<\widetilde{l}<1$),
the number of adatoms subliming from the terraces is limited by the
sublimation rate. In the sublimation limited(SL) regime, the terrace
widths in the pulse evolve as described by the equation\begin{equation}
\frac{d\widetilde{l}_{n}}{d\widetilde{t}}=\frac{(1-f)}{2}\left[\widetilde{l}_{n+1}-\widetilde{l}_{n-1}\right]\label{eq:SL}\end{equation}
which preserves any fluctuation in terrace widths and propagates it
down the crystal layers as annealing proceeds\footnote{The diffusion limited 
regime, when $\widetilde{l}\gg 1$, is omitted for simplification.  In this regime, 
$d\widetilde{l}_{n}/d\widetilde{t}=0$}.

The main insight provided by these approximations is that the dynamics governing the evolution of a 
system of steps depends on the range of terrace widths involved.  For closely spaced steps, step-step interaction
dominates while for very widely spaced steps, SL dynamics dominates.  ADL dynamics dominates for step spacings 
that are between these two regimes.

To understand how these different dynamics regimes
work together to produce a stable pulse in terrace width (plotted 
versus the terrace index), we should note that fluctuations in the terrace widths
in the ADL regime will increase in amplitude with time.  For a sinusoidal
fluctuation, this will result in the maximum terrace widths in the array becoming larger and
the minimum terrace widths becoming smaller with time.

Since the ADL regime is bounded by the step-step interaction and the DL regime,
this amplitude increase cannot continue.  Sooner or later, the terraces will get small enough to
enter the step-step interaction regime and prevent further growth of the fluctuation.  The interaction between ADL and
step-step interaction dynamics that result in stable pulses has also been demonstrated mathematically in other papers\cite{Sato:1995}.

The important new result in our paper is that for systems with low average step density, the ADL
dynamics can cause the maximum terrace widths to become large enough to enter the SL regime. 
The subsystem of steps in this SL 
regime will be subjected to boundary conditions set by the ADL dynamics, resulting in a 
pulse with a sinusoidal profile as shown in the solid line plots in Fig.~\ref{fig:adl_sl}.
If the constraint of the system size is such that the SL regime cannot be
reached, a pulse with a sharp peak will be formed as shown by the dotted plots in Fig.~\ref{fig:adl_sl}.
Thus, a pulse of terrace widths will always self-organize into a configuration that balances
the decay of fluctuations in the step-step interaction regime with
the amplification of fluctuations in the ADL regime.


We can analyze our experimental results to see if the reduction of
step density on the mesa tops due to ridge formation and breakdown
enabled the SL regime to be
reached for the terrace width distributions observed in our experiments. If this is the case,
a sinusoidal profile should be obtained if we plot the terrace index
versus terrace width. Rescaled plots of terrace widths observed on
8 different mesas for 6 samples annealed under different conditions
are shown in Fig.~\ref{fig:ren_tw}. The data suggest that the
dynamics correspond to the SL regime over the wide terraces;
the evidence is not totally conclusive since the step bunches are
not oriented exactly parallel to the mesa edge in most of our measurements,
and curved step bunches may behave somewhat differently. We may also have some
step redistribution during quenching that tends to even out the terrace distributions. 

From our theoretical analysis, we can infer that the annealing temperature
and the deposition flux will affect the magnitude of the terrace widths at which crossover occurs
between the different dynamics regimes; this will allow a degree of control over the number and the width of the widely
spaced terraces in the step array.  For example, lower annealing temperature
will favor ADL dynamics, resulting in fewer and wider terraces in the step
array as is indeed seen in our experimental results.

\section{Summary}

In summary, we have shown that novel widely spaced arrays of steps
can be produced through sublimation on top of fabricated mesa structures.
A key feature of the process is the initial formation of a ridge around
the top of each mesa in order to eliminate the large chemical potential
gradients associated with the edge. The presence of this ridge allows
the center of the mesa to become almost step-free. As the ridges widen
by diffusion, sublimation eventually takes over and causes the elimination
of the ridge and the intrusion of steps from the edge onto the mesa
surface. The new array of steps has a terrace width distribution that
persists for long annealing times. Computer simulations of the initial
low step density area on the mesa suggests that the widely spaced
step arrays decay very slowly in the sublimation limited regime and
evolve toward a stable terrace width distribution.  We also find that a degree of control may be exercised 
over number of terraces in the low step density region and their widths
by varying the annealing temperature and the deposition flux.
This provides a new method of fabricating controlled step arrays that can be used
as templates.  Observing the evolution of these systems of steps may also provide new
insights into the important kinetic process governing the dynamics of steps.


\ack
We acknowledge discussions with Prof. J.P. Sethna and the extensive use of the Cornell
Nanofabrication Facility for making samples.
This work is supported by NSF grant no. DMR-0109641 and by grant DMR-9632275 through the Cornell
Center for Materials Research. 

\bibliographystyle{elsart-num}
\bibliography{letter}

\end{document}